\documentclass[twocolumn,showpacs,preprintnumbers,amsmath,amssymb]{revtex4}

\usepackage{graphicx}
\usepackage{dcolumn}
\usepackage{bm}
\usepackage[utf8x]{inputenc}
\usepackage{float}

\begin{document}


\title{Energy Levels of One Dimensional Anharmonic Oscillator via Neural Networks}
\author{Halil Mutuk}
 \email{halilmutuk@gmail.com}
 \affiliation{Physics Department, Ondokuz Mayis University, 55139, Samsun, Turkey}

\begin{abstract}
In this work, we obtained energy levels  of one dimensional quartic  anharmonic oscillator by using neural network system. Quartic anharmonic oscillator is a very important tool in quantum mechanics and also in quantum field theory. Our results are in good agreement in high accuracy with the reference studies. 
\end{abstract}

\maketitle

\section{\label{sec:level}Introduction}
The theory of harmonic and anharmonic oscillators have an important applications in all areas of physics. Especially harmonic oscillator is an important analogy while modelling physical systems. Most quantum mechanical problems are tried to be solved by harmonic oscillator analogy. Many physical problems can be reduced to a harmonic oscillator problem with appropriate boundary conditions. This is because harmonic oscillator eigenvalue problem can be solved exactly, i.e. it has analytic solution. The Schrödinger equation with harmonic oscillator potential can be solved by using algebraic techniques, say using ladder operators. 

Yet there is no pure harmonic oscillator in the nature. That is why anharmonic oscillator have been the focus of attention since the early days of quantum mechanics \cite{1,2,3,4,5,6,7,8,9,10,11}. The general method to solve anharmonic oscillator problem is Rayleigh-Schrödinger perturbation theory but it is well known that the expansion of series in terms of eigenvalues, are convergent only small values for the coupling potential parameter. Moreover excited states can be problematic due to convergence problem. Overwhelming these difficulties is cumbersome: either solving a numerical difference equations \cite{12} or summation of strongly divergent perturbation series \citep{13,14}. To overwhelm these difficulties some nonperturbative methods are also used \citep{15,16}. The particular feature of all these approaches is that, they have some additional restrictions on the field of their applications.

In recent years, considerable progress has been made to develop mathematical methods for calculating eigenvalues and eigenfunctions of anharmonic oscillators. In \cite{17}, the authors presented new perspective for the anharmonic oscillator problem based on the SU(2) group method (SGM). In \cite{18}, Popescu showed how the variational method, in which a variational global parameter is used, can be combined with the finite element method for the study of the generalized anharmonic oscillator in D dimensions. Koscik and Okopinska applied power series method to compute the spectrum of the Schrödinger equation for central potential \cite{19}. Dong et al. in \cite{20},  proposed a new anharmonic oscillator and present the exact solutions of the Schrödinger equation with this oscillator by constructing new ladder operators. Sous in \cite{21}, by utilizing an appropriate ansatz to the wave function, reproduced the exact bound-state solutions of the radial Schrödinger equation to various exactly solvable sextic anharmonic oscillator and confining perturbed Coulomb models in D dimensions. Ikhdair and Sever utilized an appropriate ansatz to the wave function and they reproduced the exact bound-state solutions of the radial Schrödinger equation to various exactly solvable sextic an-harmonic oscillator and confining perturbed Coulomb models in D-dimensions \cite{22}. Çiftçi studied  quantum quartic anharmonic oscillator problem and obtained accurate eigenvalues for both small and large coupling parameters \cite{23} via asymptotic iteration method. Liverts et al. solved Schrödinger equation with a certain potential by first casting the Schrödinger equation into a nonlinear Riccati form and then solving that nonlinear equation analytically in the first iteration of the quasilinearization method (QLM) \cite{24}. In work of Nagy and Sailer \cite{25}, functional renormalization group methods formulated in the real-time formalism are applied to the O(N) symmetric quantum anharmonic oscillator, considered as a (0+1) dimensional quantum field-theoric model, in the next-to-leading order of the gradient expansion of the one- and two-particle irreducible effective action. Maiz and AlFaify applied Airy function approach to quantum anharmonic oscillator by setup to approach the real potential $V(x)$ of the anharmonic oscillator system as a piecewise linear potential $u(x)$ and to solve the Schrödinger equation of the system using the Airy function \cite{26}. Fernandez showed the application of group theory to a quantum-mechanical three-dimensional quartic anharmonic oscillator with O-h symmetry \cite{27}. 

All these methods above have their own shortcomings. For example, most of the usable approaches have no indication for the behaviour of the anharmonic oscillator wave functions in physically relevant ranges of their coordinates \cite{11}.

Artificial neural networks (ANNs) are being used since two decades for solving ordinary and partial differential equations. Traditional numerical techniques require discretization of domain by the number of finite domains or points where the solution functions are locally approximated. Thay are iterative methods and can contain computational complexity because of the repetition of iterations when the number of sampling points and dimensions of the problem increase. Compared to existing numerical techniques, ANNs provide some advantages such as, solution functions have a single independent variable regardless of the dimension of the problem and solutions are continuous over all the domain of integration.  The computational complexity does not increase remarkably when the the number of sampling points and dimensions of the problem increase.

In the present work, we consider the following anharmonic oscillator Hamiltonian:
\begin{equation}
H(x)=\frac{p^2}{2m}+x^2+\lambda x^4
\end{equation}
where $\lambda$ is a parameter. The eigenvalues of this Hamiltonian was studied in \citep{23,28} and extensively in \cite{29}. We obtained energy eigenvalues  of this anharmonic oscillator potential via artificial neural network. 

The paper is organized as follows. In Section \ref{sec2} we give methodology of artificial neural network and in Section \ref{sec3}, we give our numerical results. In Section \ref{sec4}, we discuss and conclude our work.

\section{Artificial Neural Network Formalism\label{sec2}}
\begin{figure}[H]
\includegraphics[width=3.4in]{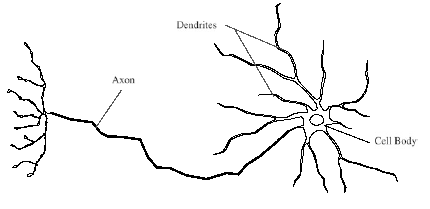}
\caption{\label{fig1} A biological neural network picture}
\end{figure}

\begin{figure}[H]
\includegraphics[width=3.4in]{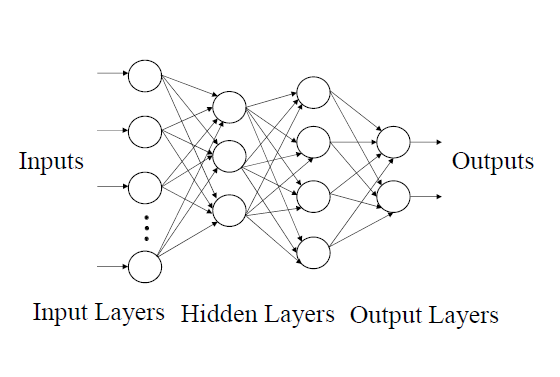}
\caption{\label{fig2} A neural network architecture}
\end{figure}

Artificial neural network systems (ANN) are simplified models of biological nervous systems. They are based on the systems of biological neurons that receive, process, and transmit information through electrical and chemical signals. In this manner artificial neural networks are made of artificial neurons and these neurons are imitations of biological neurons. A neuron has a dendrite and this can be excited by synapses of other neurons. As a result of this excitement, an electrical signal is produced and transmitted through its axon. This process repeat itself with dendrites of other neurons through its synapse. A biological neural network can be seen in Figure \ref{fig1}. 
Being a mimicked version of biological neuron, artificial neurons are connected to each other and excite each other through these connections. Neural networks are organized in layers. The input layers receive information from outside world and send them to the hidden layers. Generally, there is no process of information in input layers. A  neural network diagram is shown in Figure \ref{fig2}.

\begin{figure}[H]
\includegraphics[width=3.4in]{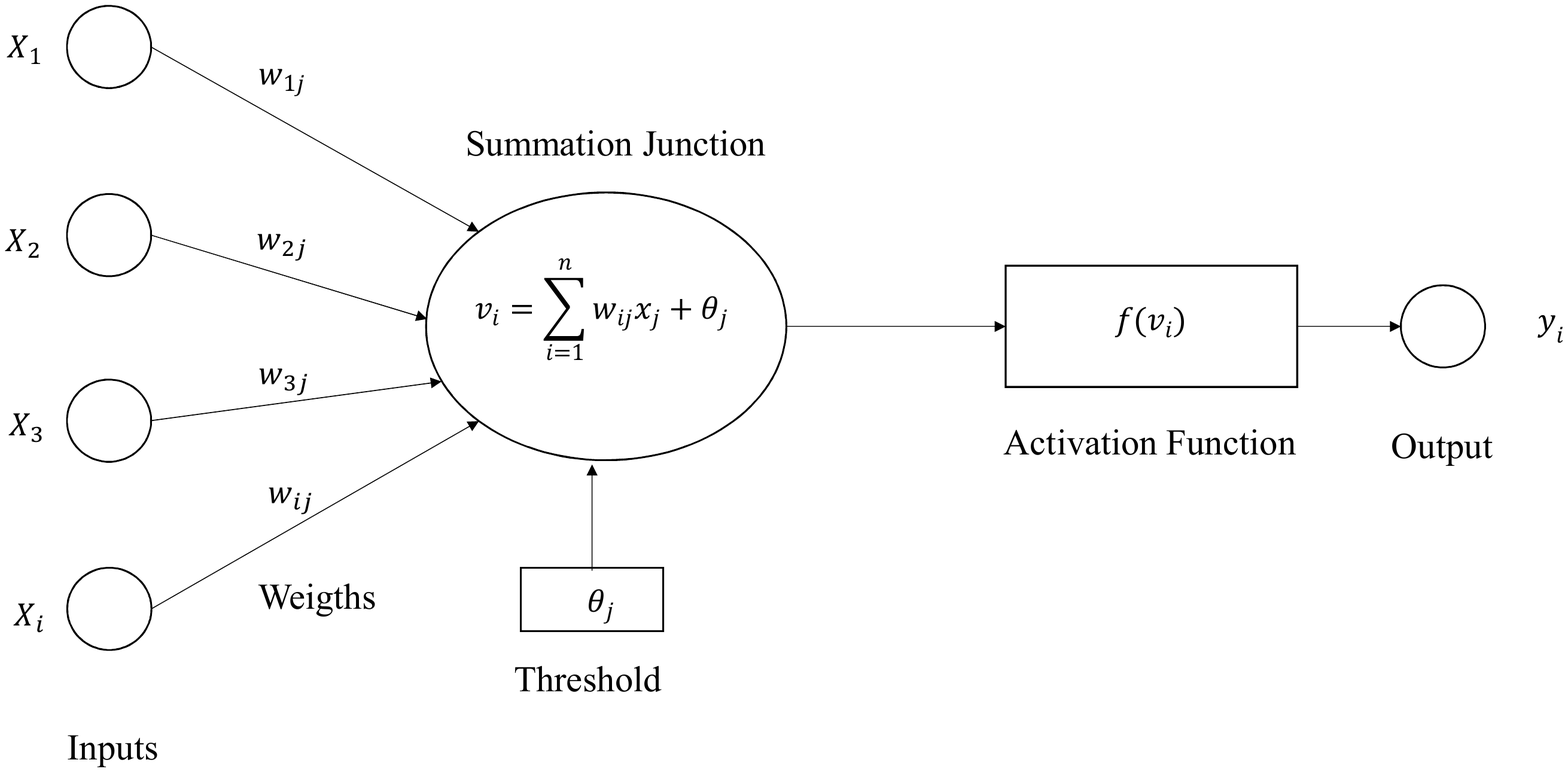}
\caption{\label{fig3} A schematic diagram of single neuron}
\end{figure}

There is one in and one out of every process element. The output is being sent to all processing elements from input layers. The hidden layers process the outputs coming from input layers and process information. They send their outputs to the output layers. Output layers produce outputs according to the inputs from input layers and send to the outside world. Every process element is connected to the all process elements from hidden layers. The mathematical formulation of a single neuron is shown in Figure \ref{fig3}. 
A neuron (perceptron in computerized systems) $N_i$ accepts a set of $n$ inputs which is an element of the set $S=\left\lbrace x_j \vert j=1,2,\cdots,n \right\rbrace$. Each input is weighted before entrance of a neuron $N_i$ by weight factor $w_{ij}$ for $j=1,2,\cdots,n$. Furthermore, it has bias term $w_0$, a threshold value $\theta_k$ which has to be reached or exceeded for the neuron to produce output signal. Mathematically, the output of the $i$-th neuron $N_i$ is
\begin{equation}
O_i=f (w_0+ \sum_{j=1}^n w_{ij}x_j).
\end{equation}
The neuron's $work$ condition can be defined as
\begin{equation}
w_0+ \sum_{j=1}^n w_{ij}x_j \geq \theta,
\end{equation}
and the input signal of the $i$-th neuron $N_i$ is
\begin{equation}
v_i=w_0+\sum_{j=1}^n w_{ij}x_j.
\end{equation}
A function $f(s)$ acts on the produced weighted signal. This function is called activation function. The output signal obtained by activation function is 
\begin{equation}
O_i=f(v_i-\theta_i).
\end{equation}
All inputs are multiplied by their weights and added together to form the net input to the neuron. This is called {\it net} and can be written as
\begin{equation}
\text{net}=\sum_{j=1} w_{ij}x_j+\theta,
\end{equation}
where $\theta$ is a threshold value which is added to the neuron. The neuron takes these inputs and produce an output $y$ by mapping function $f(net)$
\begin{equation}
y=f(\text{net})=f \left(  \sum_{j=1}^n  w_{ij}x_j+\theta \right)
\end{equation}
where $f$ is the neuron activation function. Generally the neuron output function is proposed as a threshold function but linear, sign, sigmoid or step functions are widely used. We used a sigmoid activation function 
\begin{equation}
f(x)=\frac{1}{1+e^{-x}}
\end{equation}
which is typical in multilayer perceptron neural networks. We used feed forward neural network. A feed forward neural network is the simplest form of the artifical neural network. The information moves only in one direction forward from the input nodes to the hidden and to the output nodes. If the connections between neurons are only in forward direction, the network is called a feed forward network. Connections between neurons that are in the same layer or in nonconsecutive layers are not allowed in feed forward neural networks. Figure \ref{fig2} represents a feed forward neural network.

The learning procedure of neural network in this work is unsupervised learning. In this procedure, weight adjustments are not based on comparison with some target output but some guidelines for successful learning (for example, ground state energy eigenvalue in the harmonic oscillator) is required.

\subsection{A warm up example: Harmonic oscillator}
In this subsection, we obtained ground state harmonic oscillator energy eigenvalue. The oscillator potential $V(x)=\frac{1}{2}wx^2$ in the Schrödinger equation can be solved by using algebraic method, such as ladder operators. Many problems in physics can be reduced to a harmonic oscillator problem with appropriate initial/boundary conditions. The Schrödinger equation of harmonic oscillator potential can be written with $w=m=\hbar=1$ as
\begin{equation}
-\frac{1}{2}\frac{d^2\psi(x)}{dx^2}+\frac{1}{2}x^2 \psi(x)=E \psi(x)
\end{equation}
where $E$ is the eigenvalue. The wave function satisfies the boundary conditions $\psi(-\infty)=\psi(\infty)=0$. The energy levels of the harmonic oscillator can be found by
\begin{equation}
E_n=(n+\frac{1}{2})\hbar w
\end{equation}
where $n$ is the principal quantum number and equals $n=0$ for the ground state. Once the neural network parameters (weights and biases) and a pretrain value for energy are defined, the neural network obtained the eigenvalue which is presented in Table \ref{tab:table1}. 

\begin{table}[H]
\caption{\label{tab:table1} Comparison of ground state eigenvalue of harmonic oscillator}
\begin{ruledtabular}
\begin{tabular}{ccc}
Eigenvalue& ANN & Exact\\
\hline
$E$& 0.502739612  & 0.5  \\

\end{tabular}
\end{ruledtabular}
\end{table}
Motivated from this evaluation, we obtained eigenvalues  of the anharmonic oscillator in the following section. 

\section{Numerical Results\label{sec3}}
In this section we give numerical results of the anharmonic oscillator eigenvalues. In Table \ref{tab:table2}, eigenvalues for the anharmonic oscillator are presented.

\begin{table*}
\caption{\label{tab:table2} Ground state eigenvalues for the quartic anharmonic oscillator with
different $\lambda$ values in one dimension.}
\begin{ruledtabular}
\begin{tabular}{ccccccc}
$\lambda$& $E_{\text{present}}$ & \cite{23}& \cite{28} Num. & \cite{30} \\
\hline
0.025& 1.0180097924545166  & 1.0180010006248 & 1.0180010006142 & 1.0180010006142  \\
0.05 & 1.034240512816455   & 1.0347296978234 & 1.0347296972554 & 1.0347296972530   \\
0.1  & 1.0654383929670153  & 1.0652855199    & 1.0652855096    & 1.0652854984  \\
0.2  & 1.1131419947840997  & 1.1182927141    & 1.1182926544    & 1.1182887632 \\
0.5  & 1.2430124818673798  & 1.2418546983    & 1.2418540597    & 1.2412579542  \\ 
1.0  & 1.3929635377412355  & 1.3923519526    & 1.3923516416    & 1.3853951994  \\
4.0  & 1.9029500973873372  & 1.9031372697    & 1.9031369454    & 1.769229616 \\
100.0& 4.9974247519017035  & 4.9991429       & 4.9994175452    & 4.9580018282 \\
400.0& 7.862963023360612   & 7.8620150257    & 7.8618626782    & 7.670735377 \\
2000 & 13.382962917034375  & 13.388719667 & 13.388441701 & 12.7473822552 \\ 
40000& 36.2329683376409 & 36.275234713 & 36.274458146 & 33.30734404  \\
2 $\times$ 10$^{6}$& 133.6329668678526  & 133.6029981 & 133.6001252$^{35}$ &120.199459212
\end{tabular}
\end{ruledtabular}
\end{table*}

It can be seen from Table \ref{tab:table2} that obtained eigenvalues for different $\lambda$ values are in good agreement with the given references except than the results of Ref. \cite{30}, especially for large $\lambda$ values. Note that the results of Ref. \cite{30} are multiplied by 2 for correspondence. 

\begin{table}[H]
\caption{\label{tab:table3} First eight eigenvalues of the quartic anharmonic oscillator for $\lambda=0.1$  in one dimension}
\begin{ruledtabular}
\begin{tabular}{cccccccc}
$n$& $E_{\text{present}}$ & \cite{23}& \cite{31} \\
\hline
0& 1.0654383929670153  & 1.065286 & 1.065286  \\
1& 3.3019057811028074  & 3.306872 & 3.306872   \\
2& 5.749852350602235  & 5.747959 & 5.747959   \\
3& 8.358116291676646  & 8.352686 & 8.352678  \\
4& 11.09861795976142  & 11.09837 & 11.09860  \\ 
5& 13.968331266291228  & 13.96890 & 13.96993  \\
6& 16.95956005318641  & 16.95307 & 16.95479 \\
7& 20.000623087595994  & 20.00854 & 20.04386  \\
\end{tabular}
\end{ruledtabular}
\end{table}
In Table \ref{tab:table3}, first eight eigenvalues of the quartic anharmonic oscillator are presented. Obtained eigenvalues are in good agreement with the reference studies.

\section{Discussion and Conclusion \label{sec4}}
Since two decades, machine learning technologies have been steadily developing and obtaining remarkable successes. Especially in particle physics and cosmology, large datasets are present and the current techniques can be insufficient to deal with them. Besides that, there are high energy physics researches by the neural networks \cite{32,33,34}.

The quartic anharmonic oscillator is a very important tool in quantum mechanics and also in quantum field theory. The importance of this anharmonic oscillator potential lyes in nuclear structure, quantum chemistry and quark confinement \cite{5}.

In this paper we have applied artificial neural network method to the quartic anharmonic oscillator in one dimension and have been able to find the energies. The results are in very good agreement with accurate numerical values in the literature. Our approach yielded highly accurate results for the energies of the ground state and of some excited states for this anharmonic oscillator.

\end{document}